# GALAXY WINDS
*in the age of*
# HYPERDIMENSIONAL ASTROPHYSICS

## AN ASTRO2020 SCIENCE WHITEPAPER

**Thematic Area**
Galaxy Evolution

**Principal Author**

**Grant R. Tremblay**
Center for Astrophysics | Harvard & Smithsonian

grant.tremblay@cfa.harvard.edu  |  +1 617.496.7919

**Co-Authors**

Evan Schneider[1], Alexey Vikhlinin[2], Lars Hernquist[2], Mateusz Ruszkowski[4], Benjamin Oppenheimer[5], Ralph Kraft[2], John ZuHone[2], Michael McDonald[6], Massimo Gaspari[1], Megan Donahue[7], & G. Mark Voit[7]

[1]Princeton University, [2]Center for Astrophysics | Harvard & Smithsonian, [3]University of Maryland, [4]University of Michigan, [5]University of Colorado, [6]Massachusetts Institute of Technology, [7]Michigan State University

> *"How do baryons cycle in and out of galaxies, and what do they do while they are there?"*
>
> . . .
>
> *"How do black holes grow, radiate, and influence their surroundings?"*
>
> - *New Worlds, New Horizons in Astronomy and Astrophysics*
> Two of the four Science Frontier questions posed by the Astro2010 *Galaxies Across Cosmic Time* Panel

The past decade began with the first light of ALMA and will end at the start of the new era of hyperdimensional astrophysics. Our community-wide movement toward highly multiwavelength and multidimensional datasets has enabled immense progress in each science frontier identified by the 2010 Decadal Survey, particularly with regard to black hole feedback and the cycle of baryons in galaxies. Facilities like ALMA and the next generation of integral field unit (IFU) spectrographs together enable mapping the physical conditions and kinematics of warm ionized and cold molecular gas in galaxies in unprecedented detail (Fig. 1). *JWST*'s launch at the start of the coming decade will push this capability to the rest-frame UV at redshifts $z > 6$, mapping the birth of stars in the first galaxies at cosmic dawn. Understanding of their subsequent evolution, however, now awaits an ability to map the processes that transform galaxies *directly*, rather than the *consequences* of those processes in isolation. In this paper, we argue that doing so requires an equivalent revolution in spatially resolved spectroscopy for the hot ($T > 10^6$ K) plasma that pervades galaxies, the atmospheres in which they reside, and the winds that are the engines of their evolution.

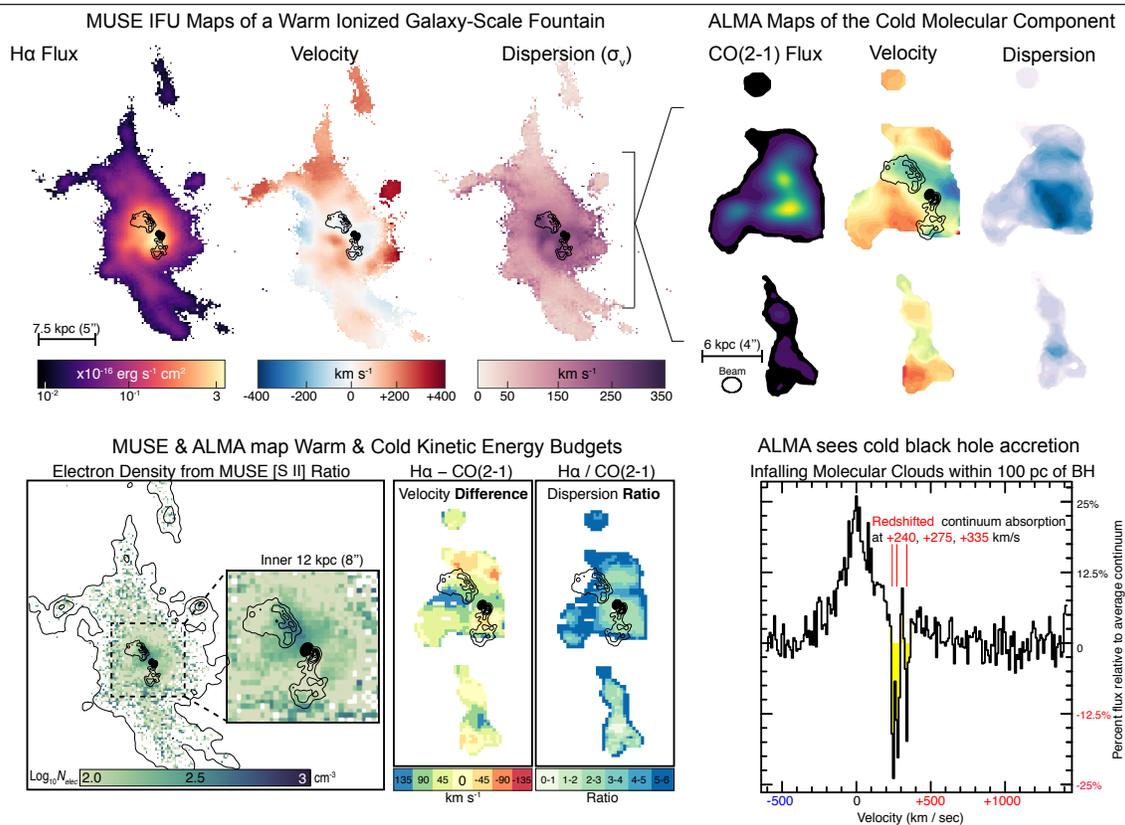

**Figure 1**. One of many recent examples of multidimensional analyses of galaxies combining datasets from facilities like ALMA and VLT/MUSE. In the above example, the combined power of both datasets reveal an outflowing and inflowing fountain of multiphase gas on galaxy-wide scales, whose morphology and kinematics are sculpted by mechanical black hole feedback. An equivalent MUSE or ALMA-like capability in the X-ray would complete the construction of spatially resolved multiphase energy budgets by mapping the kinematics of the hot ($T > 10^6$ K) gas at subarcsecond scales. Figure adapted from Tremblay *et al.* (2018).

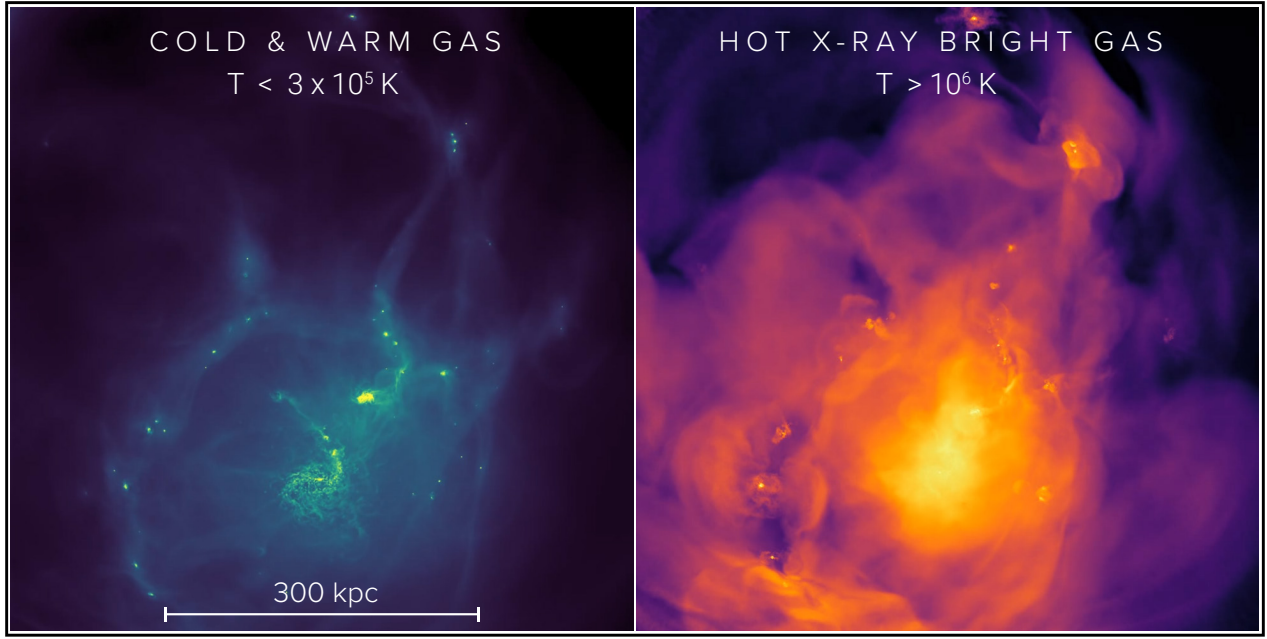

**Figure 2**. EAGLE Super-HighRes zoom simulation of the virial radius surrounding a newly formed $L^\star$ (Milky Way-mass) galaxy. The left panel shows the cold and warm component (SPH particles colder than $T \sim 10^5$ K), while the right panel shows the equivalent X-ray view of the hot ($T > 10^6$ K) gas, whose morphology and kinematics encode a history of the processes that drive the galaxy's evolution. Figure adapted from simulations courtesy of Oppenheimer and collaborators (see, e.g., the Astro2020 Science Whitepaper by Benjamin Oppenheimer *et al.*).

## THE HOT PHASE OF GALAXY WINDS

Galaxies grow as stars form amid dense clouds of cold molecular gas, and age toward quiescence when this gas is disrupted, expelled from the galaxy, or prevented from forming in the first place (see, e.g., the review by Somerville & Davé 2015). At all mass scales, galaxy evolution models now routinely invoke various forms of stellar and black hole energy feedback to reconcile observations with a theory that would otherwise over-predict the size of galaxies and the star formation history of the Universe (Fabian 2012). The past decade has seen stellar and black hole feedback achieve paradigmatic status in the field of galaxy evolution, yet both remain largely a black box with regards to how, for example, stellar superwinds or AGN mechanical energy might couple to the entropy of the ambient ISM or intracluster medium at low- and high-mass scales (respectively), or how this energy deposition is tied to the fate of cold molecular gas, from which all stars are born (e.g., Veilleux, Cecil, & Bland-Hawthorn 2005; Alexander & Hickox 2012). Progress in this vital science will be reliant on a better understanding of the hot circumgalactic and intra(group/cluster) medium, which serves as a fuel reservoir powering the processes that drive them (see, e.g., the Astro2020 Science Whitepapers by Oppenheimer *et al.*, Peeples *et al.*, and Tumlinson *et al.*). However, we require direct observations of ongoing feedback to truly understand its physics.

The Astro2020 Committee will receive many excellent whitepapers with recommendations for studying feedback in the UV through sub-mm regimes. Yet in many ways, the UV and sub-mm data are equivalent to observing only the sparks in a fire. We now need the capability to observe and map the flame. **It is therefore vitally important to spatially map the kinematics and physical conditions of the hot ($T > 10^6$ K) phase of galaxy winds across a wide range in redshift and across decades in the galaxy mass scale,** achieving parity with a capability that already exists for the warm ionized and cold molecular gas phases.

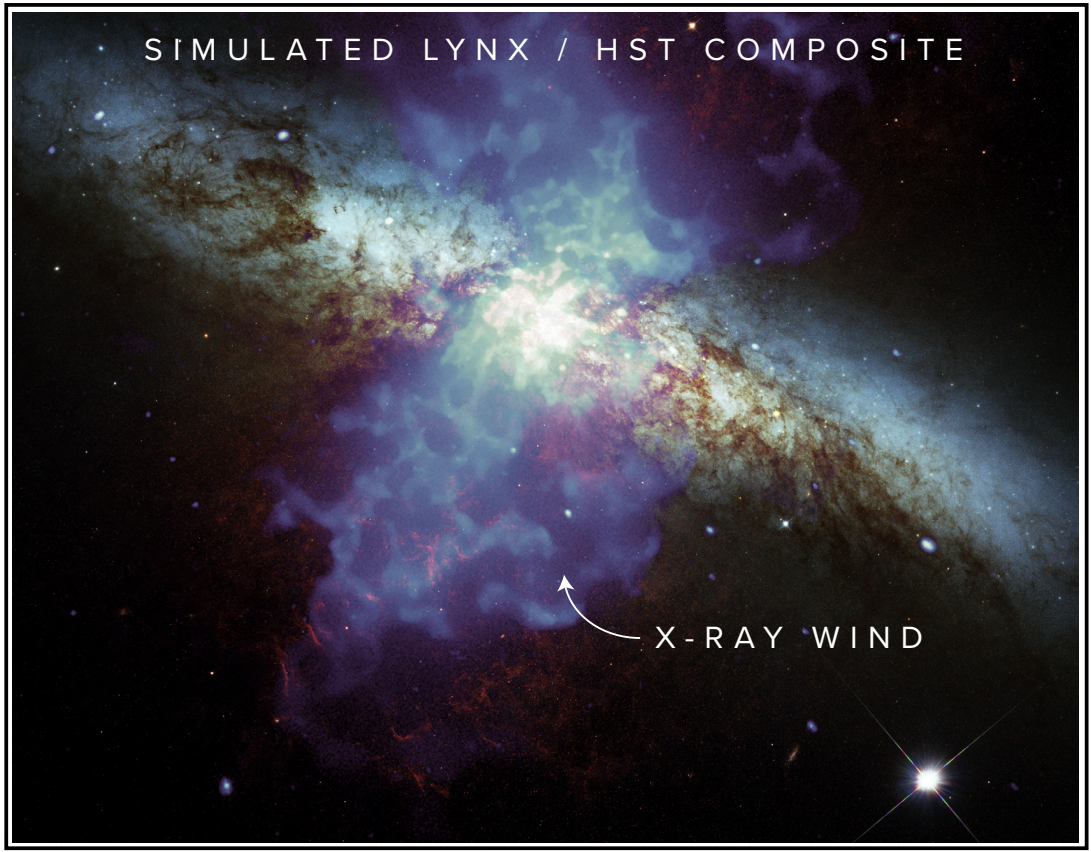
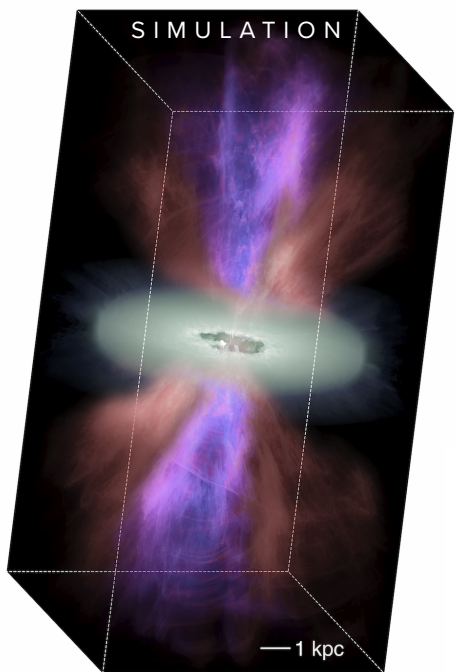
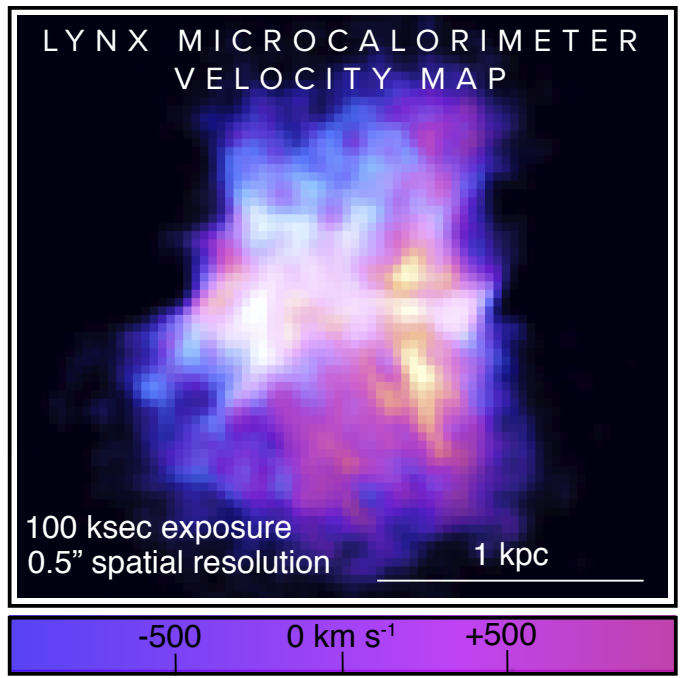

**Figure 3**. (*Top*) A mock *Lynx* / HST composite of M82, the canonical (and first-known) stellar superwind in a galaxy. (*Bottom left*) A GPU simulation, using the CHOLLA Galactic Outflow simulation suite, of an M82-like galaxy at 5 pc resolution over a scale of 20 kpc. Colors in the outflow encode its absolute velocity. Figure adapted from Schneider *et al.* (2018). (*Bottom right*) A 100 ksec mock observation of the same simulated galaxy with the *Lynx* X-ray Microcalorimeter. The instrument would be capable of resolving the line-of-sight velocity of hot gas in the outflow at ~30 km s$^{-1}$ spectral resolution on arcsecond scales.

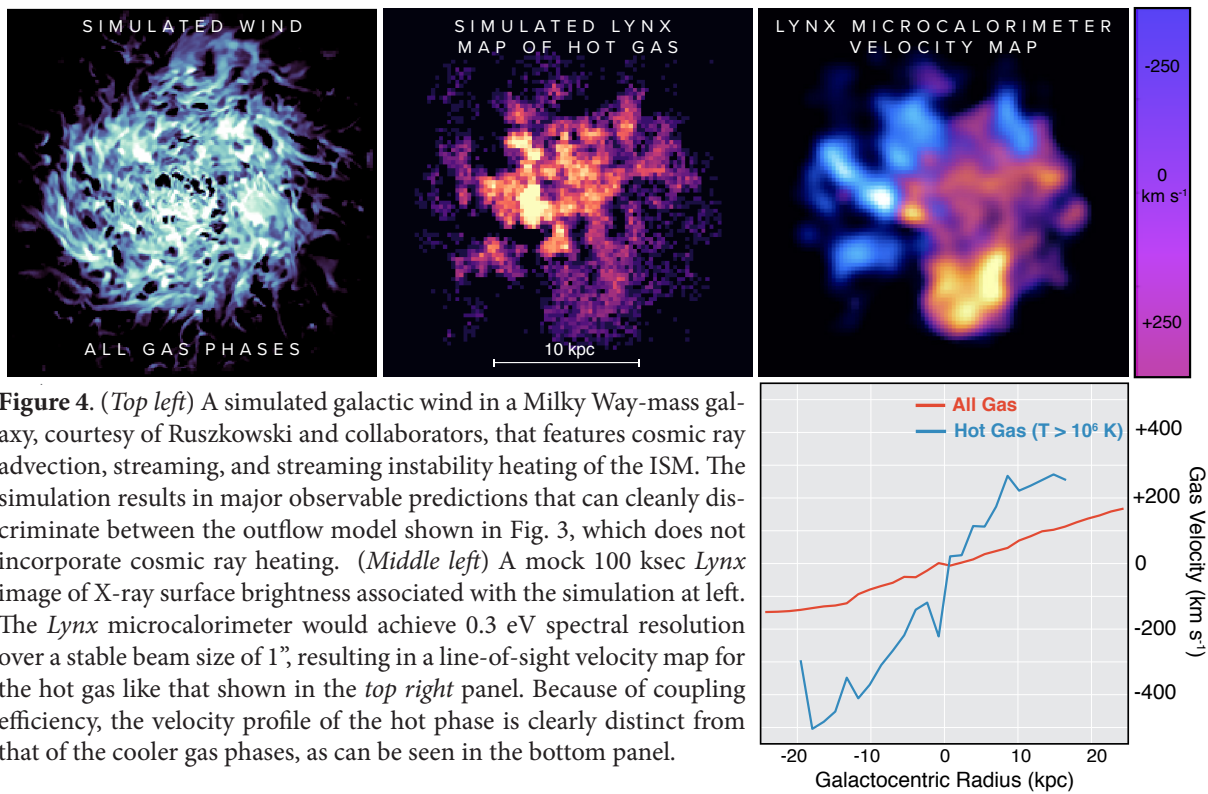

**Figure 4**. (*Top left*) A simulated galactic wind in a Milky Way-mass galaxy, courtesy of Ruszkowski and collaborators, that features cosmic ray advection, streaming, and streaming instability heating of the ISM. The simulation results in major observable predictions that can cleanly discriminate between the outflow model shown in Fig. 3, which does not incorporate cosmic ray heating. (*Middle left*) A mock 100 ksec *Lynx* image of X-ray surface brightness associated with the simulation at left. The *Lynx* microcalorimeter would achieve 0.3 eV spectral resolution over a stable beam size of 1", resulting in a line-of-sight velocity map for the hot gas like that shown in the *top right* panel. Because of coupling efficiency, the velocity profile of the hot phase is clearly distinct from that of the cooler gas phases, as can be seen in the bottom panel.

Simulations give us guidance that the structure in winds is rich and abundant, containing essential observables that are fundamental for our understanding of how these winds operate. Because those structures often subtend less than an arcsecond on the sky even for nearby objects like M82, sub-arcsecond resolution (coupled to very high soft X-ray sensitivity) is needed to enable direct comparison with data from flagship ground-based IFUs like MUSE and interferometers like ALMA at matching spatial scales. The science enabled by such a parity in capability would be entirely transformational for our understanding of galaxy evolution, and capitalize on the discoveries that (e.g.) *JWST*, ALMA, and the ELTs will make in the decades to come. Subarcsecond spatially resolved spectroscopy at ~30 km s$^{-1}$ velocity resolution in the X-ray band will enable these paradigm-shaping advancements along multiple fundamental questions relevant to both stellar and black hole feedback at all mass scales.

**One major, overarching question for galaxy evolution is: How much energy lies in galaxy winds, where is it deposited, and how does this deposition take place?** While both stellar and black hole feedback launches winds that are inherently multiphase, thermalization and coupling efficiency between the wind driver (e.g., clustered supernovae; jets launched by black holes) and the ambient gas is highest for the hot gas (e.g., Tremblay *et al.* 2018, and references therein). X-ray observations are therefore a direct observation of the bulk of the energy carried by superwinds, and a probe of the most efficient mechanism by which metal-rich gas is transported on galaxy-wide scales and beyond. In addition to the energy budget, these data are also crucial for understanding the physics of wind launching. As an example, we compare the structure of winds launched with and without a significant cosmic ray component in Figs. 3 and 4.

The observable discriminants between these two wind drivers are largely found in the velocity structure, clumpiness of the hot plasma, and how it couples to the cold gas. These signals are mappable only by a subarcscond, high spectral resolution X-ray microcalorimeter in space.

Such an instrument can also measure the shock strength (and thus wind velocity) at the boundary of the wind and cooler phases if the ISM/CGM. Multiple spatial and spectral resolution elements would be needed across the wind cone to map (e.g.) the iron lines that trace turbulent broadening in the wind fluid (e.g., Doppler *b*), as well as its the non-equilibrium ionization. For individual galaxies, shocks heat gas within a few kpc of the nucleus, such that only a subarcsecond X-ray telescope can measure the shock strength on scales associated with the nuclear regions of galaxies beyond Virgo. The issue of spatial resolution is similarly important in measurements of the thermalization efficiency across the galaxy mass scale and throughout galaxy disks, as well as the mass loading factor in winds and fountains as a function of galactocentric radius.

The most energetic examples of black hole feedback are found in galaxy clusters (reviews by, e.g., McNamara & Nulsen 2007; Fabian 2012), which also serve as astrophysical laboratories for studies of plasma physics effects coupled to turbulent energy dissipation from feedback (e.g., Zhuravleva *et al.* 2014). *Athena* will be transformational for feedback studies in clusters. An arcsecond-resolution X-ray microcalorimeter will be key for studies of associated plasma physics (see, the Astro2020 whitepaper by Ruszkowski *et al.*).

## REQUIREMENTS

The advances described in this whitepaper depend on the advent of a space-based X-ray microcalorimeter with subarcsecond spatial resolution, as well as the ability to centroid velocities to ~30 km s$^{-1}$ in the soft (~0.2 - 1 keV) band, allowing for mapping of the ~100 km s$^{-1}$ velocities expected in galactic winds. The required raw resolving power of the microcalorimeter is $R \sim 2000$ at 0.6 keV. *XRISM* and the X-IFU instrument aboard *Athena* will miss both the spectral and angular resolution requirements by more than an order of magnitude[1], as their science goals are largely orthogonal to those outlined in this paper. The realization of our vision requires further advances in microcalorimeter focal plane and readout technologies, such as those considered for the *Lynx* X-ray Microcalorimeter (see, e.g., the *Lynx* Interim Report).

---

[1] *Athena* achieves similar resolving powers for the 6.5 keV Iron line, but not for the soft band.